\documentclass{article}
\usepackage{spconf,amsmath,graphicx}
\usepackage[table]{xcolor}
\usepackage{booktabs}
\usepackage{hhline}
\usepackage{bbm}
\usepackage[tableposition=top]{caption}

\DeclareMathOperator*{\minn}{min}
\title{Generative Adversarial Speaker Embedding Networks for Domain Robust End-to-End Speaker Verification}
\name{Gautam Bhattacharya$^{1,2}$, Joao Monteiro$^{2}$, Jahangir Alam$^{2}$, Patrick Kenny$^{2}$}
\address{ $^1$McGill University \\
          $^2$Computer Research Institute of Montreal}
\begin{document}
%
\maketitle
\begin{abstract}
This article presents a novel approach for learning domain-invariant speaker embeddings using Generative Adversarial Networks. The main idea is to confuse a domain discriminator so that is can't tell if embeddings are from the source or target domains. We train several GAN variants using our proposed framework and apply them to the speaker verification task. On the challenging NIST-SRE 2016 dataset, we are able to match the performance of a strong baseline x-vector system. In contrast to the the baseline systems which are dependent on dimensionality reduction (LDA) and an external classifier (PLDA), our proposed speaker embeddings can be scored using simple cosine distance. This is achieved by optimizing our models end-to-end, using an angular margin loss function. Furthermore, we are able to significantly boost verification performance by averaging our different GAN models at the score level, achieving a relative improvement of 7.2\% over the baseline.

\end{abstract}
\section{Introduction}
\label{sec:intro}


\vspace{0.5em} \textit{Speaker Embeddings} are low-dimensional vector representations that contain information relevant to a person's identity. Unlike the first generation of speaker embeddings \cite{p1}, Deep Neural Networks (DNN) are powerful are able to learn robust, distributed representations directly from data. While neural network based speaker embeddings are finding application in several downstream tasks such as speech recognition, synthesis and source separation \cite{p2,p3}, they have most widely been applied to speaker verification \cite{p4,p5,p6,p7}. 

\noindent In the simplest case, we are given two recordings from which we can obtain the corresponding speaker embeddings. We can then obtain a verification score by computing the cosine distance (or some other simple metric) between the two embeddings. For such models to be robust, they typically need to optimize the distance metric directly. Such models are often referred to as - end-to-end, and in the context of speaker verification have proved challenging to train. Consequently, state-of-the-art DNN speaker embeddings employ the same recipe as the popular i-vector model \cite{p8} and use LDA for dimensionality reduction and PLDA for scoring verification trials.

\vspace{0.5em}\noindent  Verifying a speaker's identity is a challenging problem. Modern speaker verification datasets like NIST-SRE 2016 add to this challenge by introducing a mismatch between the distributions of the training and test data \cite{p9}. This phenomena is referred to as domain or covariate shift. In the case of NIST-SRE 2016, the test data consists of Cantonese and Tagalog speakers, whereas the vast majority of training speakers are talking in English. 

\noindent NIST also provide a small amount of unlabelled, \textit{in-domain, target} data, that can be used to compensate for the domain shift. Most the domain adaptation techniques that have been proposed for speaker verification have been proposed on top of i-vectors or x-vectors. In our recent work, we proposed to learn domain-invariant speaker embeddings using domain adversarial training \cite{p10}. The proposed DANSE model is also optimized end-to-end, and produces competitive verification performance using simple cosine scoring.

\vspace{0.5em}\noindent In this article we extend our previous work by using Generative Adversarial Networks (GAN) to achieve unsupervised domain adaptation/invariance. We are motivated to explore this line of research given the success of GAN-based domain adaptation methods in computer vision \cite{p11,p12}. The DANSE model uses gradient reversal to achieve domain invariance, and optimizes the true minmax objective of the adversarial game \cite{p13}. Replacing gradient reversal with an explicit GAN game offers the following advantages. Firstly, GAN optimization based on an inverted label loss provides stronger gradients for learning invariant mappings than gradient reversal. Secondly, the GAN framework is more general and extendable than gradient reversal. From our analysis we show that different GAN variants produce different transformations of the feature space, and our experiments indicate that combining these feature spaces is beneficial in terms of verification performance. Both methods cast domain adaptation/invariance as an adversarial game - generate features or embeddings such that a discriminator cannot tell if they come from the source or target domain. 
 

The nature of the adversarial game makes training GAN models inherently challenging \cite{p23}. We found that a simple way to stabilize the training of our models was to make the GANs conditional. Specifically we propose to use a modified version of the Auxiliary Classifier GAN (AuxGAN)\cite{p15}. We were able to train several GAN variants within our proposed framework, both with and without the auxiliary classifier stabilization. We find that all of the GAN models outperform the DANSE model in terms of verification performance. All the adversarial speaker embeddings proposed in this work outperform our i-vector baseline by a large margin, but are not quite able to match a state-of-the-art x-vector baseline. Interestingly, we find that we are able to beat this baseline by simply averaging the scores (cosine distances) of our different GAN models.



 
\section{Domain Adaptation with GAN\MakeLowercase{s}}
A GAN is trained through a minimax game between a generator, that maps noise vectors in the image space, and a discriminator, trained  to  discriminate  generated  images  from  real ones. This structure also makes GANs ideal candidates for unsupervised domain adaptation - wherein the adversarial game takes place in vector or embedding space, and the discriminators goal is to differentiate between \textit{embeddings} from the source and target domains \cite{p11,p12}. 

\noindent The key component of these models is the domain discriminator (red block in Fig. 1). From our experiments we find that different discriminator configurations corresponding to several GAN variants lead to different transformations of feature space. In a standard or vanilla GAN, the discriminator is trained by optimizing the Binary Cross-Entropy Loss (BCE). The GAN game is represented by the following general form:


\begin{equation}
\begin{split}
    & \minn_{D} \ \mathcal{L}_{adv_{D}}(\textbf{X}_{s},\textbf{X}_{t},E) \\
    & \min_{E} \ \mathcal{L}_{adv_{E}}(\textbf{X}_{s},\textbf{X}_{t},D) \\
\end{split}
\end{equation}

Where $D$ is the domain discriminator, $E$ is the embedding function (generator) and $\textbf{X}_{s}$, $\textbf{X}_{t}$ represent the source and target data respectively. When training the generator, the typical approach is to use the BCE loss with inverted labels. This splits the optimization into two independent objectives, one for the generator and one for the discriminator. 

\vspace{0.5em}\noindent Eq. 1 represents the most general form of the adversarial objective, and several adversarial domain adaptation techniques are encapsulated by this framework. By setting $\mathcal{L}_{adv_{E}} = - \mathcal{L}_{adv_{D}}$, we obtain the gradient reversal model. The advantage of using the GAN objective is that it has the same fixed-point properties as the gradient reversal but provides stronger gradients to the target mapping \cite{p11}.


\section{Generative Adversarial Speaker Embedding Networks}

Our goal is encourage a speaker embedding model to learn domain invariant features by playing a GAN game between the feature extractor (generator) and a domain discriminator.
Recently we proposed the DANSE model for speaker verification, which integrates adversarial training with a speaker embedding model. We were able to show that by combining verification with domain adversarial training, we are able to achieve competitive performance using simple cosine scoring. In this work we choose to replace the gradient reversal method with an explicit GAN game. We show that the framework is robust, with several GAN variants displaying good speaker verification performance. 

\begin{figure}[htb]
  \includegraphics[scale=0.75]{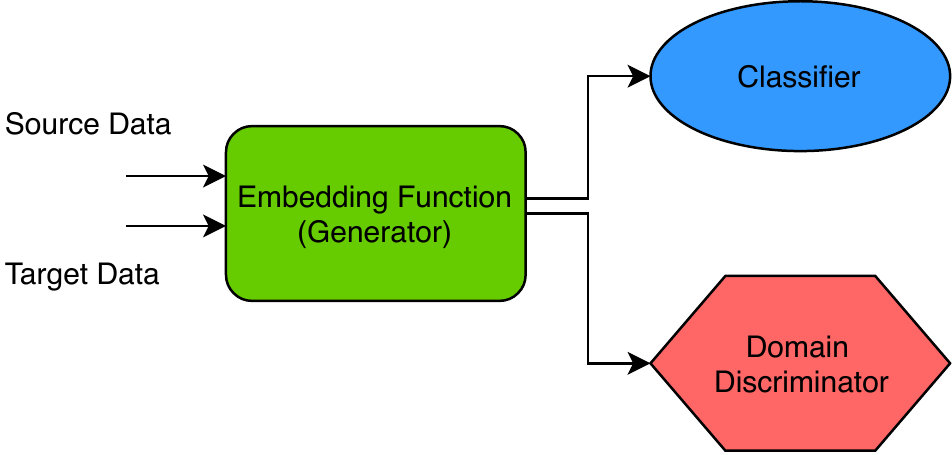}
  \label{fig:danse}
\end{figure}

\vspace{0.5em}\noindent While the GAN game encourages domain invariance, we also require the embeddings to be speaker discriminative. These characteristics are induced by minimizing an appropriate task loss:


\begin{equation}
\begin{split}
    &\minn_{E,C} \mathcal{L}_{task}(\textbf{X}_{s},\textbf{Y}_{s}) = \\  
    &-\mathbbm{E}_{(x_{s},y_{s})\sim (X_{s},Y_{s})} \sum_{k=1}^{K} \mathbbm{1}_{[k=y_{s}]} \log(C(E^{s}(x_{s}))
\end{split}
\end{equation}

Where $C(.)$ is a classifier and $E(.)$ is the Embedding function. We use the Additive Margin Softmax loss for $L_{task}$, which directly optimizes the optimizes for cosine similarity between classes \cite{p16}. It offers a clear advantage over standard cross-entropy training \cite{p10}. The AM-Softmax loss maintains the same structural form as the cross-entropy loss:

\begin{equation}
\begin{split}
    \mathcal{L}_{task} & = - \frac{1}{n} \sum^{n}_{i=1} \log \frac{e^{s.(cos\theta_{y_{i}} - m)}}{e^{cos\theta_{s.(y_{i}} - m)} + \sum_{j\neq y_{i}} e^{s.(cos\theta_{j})} } \\
    & = - \frac{1}{n}\sum^{n}_{i=1}\log \frac{e^{s.(W^{T}\boldsymbol{f_{i}} - m)}}{e^{s.(W^{T}\boldsymbol{f_{i}} - m)} + \sum_{j\neq y_{i}} e^{s.(W^{T}\boldsymbol{f_{j}})}}
\end{split}
\end{equation}

\vspace{0.5em} Where $s$ is a scale factor and $m$ is a margin. During training, the first step is to update the $C(.)$ and $E(.)$ (blue and green blocks in Fig. 1) by minimizing $\mathcal{L}_{task}$. Next we extract source and target embeddings for GAN training. We first update the domain discriminator $D(.)$ using BCE loss on both source and target data:

\begin{equation}
\begin{split}
        &\mathcal{L}_{adv_{D}}(\textbf{X}_{s},\textbf{X}_{t},E) = \\ \ \ \ \ \ \ \ &-\mathbbm{E}_{x_{s}\sim X_{s}} [\log(D(E(x_{s})]  \\
       \ \ \ \ \ \ \ \ \ \  &-\mathbbm{E}_{x_{t}\sim X_{t}} [\log(1 - D(E(x_{t})]
\end{split}
\end{equation}

\noindent Finally, we update the generator/embedding function, $E(.)$ to fool the discriminator using the inverted label loss:
\begin{equation}
    \mathcal{L}_{adv_{E}}(\textbf{X}_{s},\textbf{X}_{t},D) =  -\mathbbm{E}_{x_{s}\sim X_{s}} [\log(D(E(x_{t})]
\end{equation}

\noindent Note that the embedding function $E(.)$ gets updated twice, first with the task loss followed by the adversarial loss. 






\subsection{Auxiliary Classifier GAN}

The Auxiliary Classifier GAN (AuxGAN) model augments the standard GAN framework with an auxiliary loss to perform conditional image generation \cite{p15}. This approach aims to predict side information (such as class labels), as opposed to feeding the same information to the generator and discriminator. In the context of this work, we use the auxiliary loss for regularization and representation learning.

\begin{equation}
\begin{split}
    & \minn_{D} \ \mathcal{L}_{adv_{D}}(\textbf{X}_{s},\textbf{X}_{t},E) + \mathcal{L}_{Aux}(\textbf{X}_{s},Y_{s}) \\
    &  \minn_{E} \ \mathcal{L}_{adv_{E}}(\textbf{X}_{s},\textbf{X}_{t},D) + \mathcal{L}_{Aux}(\textbf{X}_{s},Y_{s}) \\
\end{split}
\end{equation}

Eq. (6) is a modified version of the AuxGAN objective. In particular, the original formulation also uses the auxiliary loss to train on fake data as well (with fake data being assigned its own unique label). We use the cross-entropy loss over speakers as the auxiliary loss. In our experiments we found that using an auxiliary classifier helps stabilize all the GAN variants, but does not always lead to the better verification performance. 


\subsection{GAN Variants}

Since their introduction, GANs have been one of the most researched topics in the deep learning community. Several variations of the original formulation have been proposed, each with different generative characteristics and stability issues. In this work we explore two GAN variants in addition to the standard GAN - Least-Squares GAN and Reletavistic GAN \cite{p17,p18}. These models differ in the structure of the discriminator network. We show that each variant transforms the feature space in different ways, with all the model showing mostly similar performance. Additionally we see that by fusing the performance of all GAN variants together through score averaging we achieve the best overall performance.

\section{Experiments \& Results}
\label{sec:page}


\vspace{0.5em}\noindent \textbf{Training Data (Source):} We used audio from previous NIST-SRE evaluations (2004-2010) and Switchboard Cellular audio for training the proposed DANSE model as well as the x-vector and i-vector baseline systems. We also augment our data with noise and reverberation, as in \cite{p4}. We add 128k noisy copies to the clean speech, ending up with 220k recordings in our training set. For training adversarial models we filter out speakers with less than 5 recordings, ending up with approximately 6000 speakers. Whereas the x-vector and i-vector systems were trained using the Kaldi recipe. We note that the vast majority of our training data consists of English speakers, and is recorded over telephone/cellular channels.

\vspace{0.5em}\noindent \textbf{Model:} In order to make a fair comparison, we use an identical network to the DANSE model. The \textit{Embedding function/Generator}, $E(.)$, consists of a $3X23$ input convolutional layer, 4 residual blocks [3,4,6,3], an attentive statistics layer and two fully connected layers (512,512). The \textit{classifier}, $C(.)$, module consists of a fully connected layer (64)  and the AM-Softmax output layer. The former is the final domain invariant speaker embedding extracted during evaluation. Finally, the \textit{domain discriminator} module consists of two fully connected layers (256,256) and a binary cross-entropy output layer. Exponential Linear Units (ELU) are used as non-linear activations for all layers of the network. Batch Normalization is used on all layers expect the attentive statistics layer. The $s$ and $m$ hyper-parameters of the AM-Softmax loss are set to 30 and 0.6 respectively.We refer the reader to \cite{p10} for a detailed description of the model.

\vspace{0.5em}{\noindent} \textbf{Optimization:} We start by pre-training the Embedding function using standard cross-entropy training. Pre-training is carried out using the RMSprop optimizer with a learning rate (lr) of $0.001$. For training GAN based speaker embedding models we use different optimizers for training the three networks (Embedding function,Classifier, Discriminator). The classifier is optimized using RMSprop with $lr$=$0.003$, while the domain classifier and feature extractor are trained using SGD with $lr$=$0.001$. We were able to train all our GAN models using the same set of hyper-parameters. We used performance on held out validation set to determine when to stop training.

\vspace{0.5em}{\noindent} \textbf{Data Sampling:} We use an extremely simple approach for sampling data during training. We sample random chunks of audio (3-8 seconds) from each recording in the training set. We sample each recording 10 times to define an epoch. For each mini-batch of source data, we randomly sample (with repetition) a mini-batch from the unlabelled adaptation data for GAN training.

\begin{figure*}[htb]
  \includegraphics[width=\textwidth,height=6cm,scale=0.75]{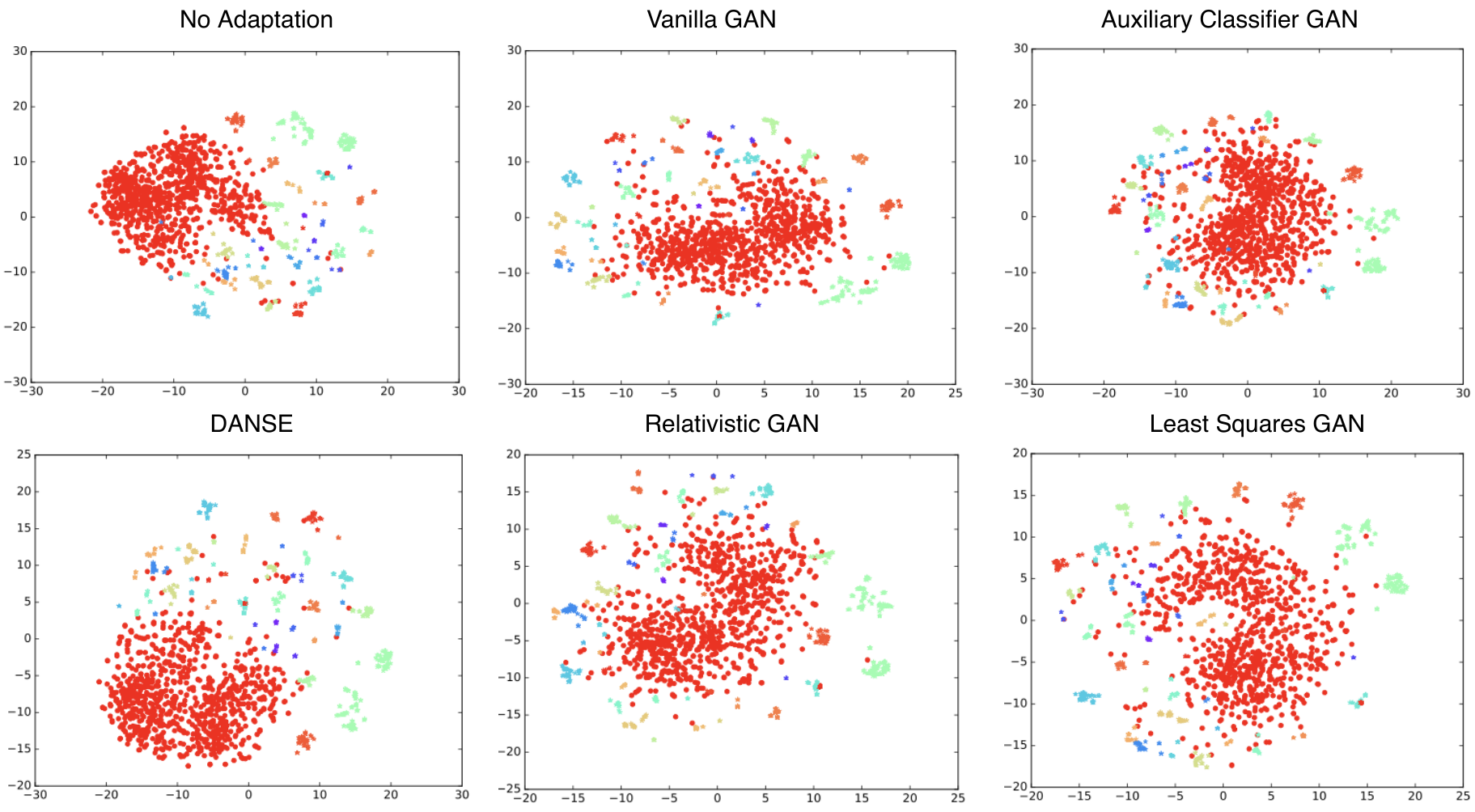}
  \caption{t-SNE Visualization \cite{p24} of Embedding Space. Large red cluster represents target data. Other colours are source domain speakers.}
  \label{fig:danse1}

\end{figure*}

\vspace{0.5em}\noindent \textbf{Speaker Verification:} At test time we discard the domain discriminator branch of the model, as it is not needed for extracting embeddings. Extraction is done by performing a forward pass on the full recording, and using the 64-dimensional $fc3$ layer as our speaker embedding. Verification trials are scored using cosine distance. Verification performance is reported in terms of Equal Error Rate (EER).

\vspace{0.5em}\noindent \textbf{NIST-SRE 2016:} Unlike previous years, The 2016 edition of the NIST-SRE introduced a challenging new dataset containing Cantonese and Tagalog speakers. We use the Kaldi recipes for our baseline i-vector and x-vector systems. We note this x-vector baseline may be considered as state-of-the-art performance on this dataset. 

\noindent\textbf{Adaptation Data (Target):} 2272 unlabelled, target data recordings are provided to adapt verification systems. 

\begin{table}[htb]
\begin{center}

\begin{tabular}{*5l}    \toprule
\textbf{\emph{Model}} & \textbf{\emph{Classifier}} &\textbf{\emph{Cantonese}}& \textbf{\emph{Tagalog}}& \textbf{\emph{Pooled}}  \\\midrule
i-vector    & LDA/PLDA  & 9.51  & 17.61  & 13.65  \\ \hline
x-vector & COSINE & 36.44 & 41.07 & 38.69\\
\textbf{x-vector} &\textbf{ LDA/PLDA} & \textbf{7.52} & \textbf{15.96} & \textbf{11.73}\\ 
x-vector & PLDA & 7.99 & 18.46  & 13.32\\ \bottomrule

 \hline
\end{tabular}
\caption{Baseline Systems}

\end{center}
\end{table}

\vspace{-1cm}

\begin{table}[htb]
\begin{center}
\begin{tabular}{*5l}    \toprule
\textbf{\emph{Model}} & \textbf{\emph{Classifier}} & \textbf{\emph{Cantonese}}& \textbf{\emph{Tagalog}}& \textbf{\emph{Pooled}}  \\\midrule
DANSE & COSINE & 8.84  & 17.87 & 13.29\\
SGAN    & COSINE  & 8.32  & 17.51  & 12.93 \\ 
AuxGAN & COSINE & 7.96 & 15.90 & 11.93\\
\rowcolor{green}
\textbf{LSGAN} & \textbf{COSINE} &  \textbf{7.90} & \textbf{15.63} & \textbf{11.74}\\ 
RelGAN & COSINE & 8.01 & 16.22  & 12.21\\ 
\rowcolor{cyan}
\textbf{FuseGAN} & \textbf{COSINE} & \textbf{6.93}  & \textbf{14.84} & \textbf{10.88}\\\bottomrule
 \hline
\end{tabular}
\caption{\textbf{ DANSE:} Gradient Reversal, \textbf{SGAN:} standard, \textbf{AuxGAN:} auxiliary classifier, \textbf{LSGAN:} least squares, \textbf{RelGAN:} relativistic, \textbf{FuseGAN:} score averaging}
\end{center}
\end{table}

\vspace{-1cm}

\vspace{0.5em}\noindent Tables 1. \& 2. compare the performance of the proposed adversarial speaker embeddings against the baseline systems. Among the baseline systems we see that the DNN based x-vector system clearly outperforms i-vectors, however only with the addition of LDA dimensionality reduction.

\vspace{0.5em}\noindent We see that all the GAN based models outperform DANSE by a large margin. This result is confirms the hypothesis that the GAN game provides a more effective learning signal than the gradient reversal framework. All the individual GAN models also outperform the i-vector baseline by a large margin, and the best model (LSGAN) is able to match the x-vector baseline. We find that we are able to significantly improve on this this baseline by simply averaging the verification scores (cosine distances) of the AuxGAN, LSGAN and RelGAN embeddings. The FuseGAN ensemble achieves the lowest EER on all three splits of the verification trials. 

\vspace{0.5em}\noindent This result is interesting as it suggests that the different GAN models cover different modes of the target data distribution, and catpure speaker information that is different yet complimentary. Comparing the transformed feature spaces in Fig. 2, we see that the DANSE model appears to mainly rotate the feature space, whereas the transformations induced by the GAN models is more pronounced. Importantly we see that the source data speaker clusters remain intact after adversarial adaptation. 

\section{Conclusion}

We presented a novel framework for learning domain invariant speaker embeddings using GANs. By learning several different GAN variants and combining them at the score level, we are able to achieve state-of-the-art verification performance. Additionally our models are optimized end-to-end and can be scored using simple cosine scoring. In future work we will consider other adversarial strategies like combining feature space and data space GANs \cite{p19,p20} and GAN-based feature space augmentation methods \cite{p21}.

\bibliographystyle{IEEEtran}
\bibliography{mybib}

\end{document}